\documentclass[aps,prc,twocolumn,superscriptaddress]{revtex4}
\usepackage{graphicx} 
\begin{document}
\bibliographystyle{apsrev}


\title{Auxiliary Field Diffusion Monte Carlo calculation of
ground state properties of neutron drops}




\author{F. Pederiva}
\email{pederiva@science.unitn.it}
\affiliation{Dipartimento di Fisica dell'Universit\'{a} di Trento,
and \\ INFM {\sl DEMOCRITOS} National Simulation Center \\
I-38050 Povo, Trento, Italy}
\author{A. Sarsa}
\email{fa1sarua@uco.es}
\affiliation{International School for Advanced Studies, SISSA,
and \\ INFM {\sl DEMOCRITOS} National Simulation Center \\ Via Beirut
I-34014 Trieste, Italy}
\affiliation{
Departamento de Fisica, Universidad de Cordoba, Spain}
\author{K. E. Schmidt}
\email{kevin.schmidt@asu.edu}
\affiliation{International School for Advanced Studies, SISSA,
and \\ INFM {\sl DEMOCRITOS} National Simulation Center \\ Via Beirut
I-34014 Trieste, Italy}
\affiliation{Department of Physics and Astronomy,
Arizona State University, Tempe, AZ, 85287}
\author{S. Fantoni}
\email{fantoni@sissa.it}
\affiliation{International School for Advanced Studies, SISSA,
and \\ INFM {\sl DEMOCRITOS} National Simulation Center \\ Via Beirut
I-34014 Trieste, Italy}


\date{\today}

\begin{abstract}
The Auxiliary Field Diffusion Monte Carlo 
method has been applied to simulate droplets of 7 and 
8 neutrons. Results for realistic nucleon-nucleon
interactions, which include
tensor, spin--orbit and three--body forces, 
plus a standard one--body confining 
potential, have been compared with analogous calculations obtained with
Green's Function Monte Carlo methods. 
We have studied the dependence of the binding energy, the one--body 
density and the spin--orbit splittings of $^7n$ on the
depth of the confining potential.
The results obtained show an overall agreement between the two quantum Monte
Carlo methods, although there persist differences in the evaluation of
spin--orbit forces, as previously indicated by bulk neutron matter calculations.
Energy density functional models, largely used in astrophysical applications,
seem to provide results significantly different from those of 
quantum simulations. Given its scaling behavior in the number of nucleons,
the Auxiliary Field Diffusion Monte Carlo method seems to be one of the best 
candidate to perform {\sl ab initio} calculations on neutron rich nuclei. 
\end{abstract}


\maketitle


\section{Introduction}

Neutron rich matter is a subject of fundamental 
interest in nuclear astrophysics. Important phenomena, like the
structure and evolution of compact stellar objects, the $r$--process 
in nucleosynthesis or the mechanism of
supernovae explosion cannot be understood without a deep knowledge of
the properties of such matter\cite{pethick95,raffelt96,kratz93,
akmal98,lattimer01}.

In recent years the relevance of nucleon-nucleon (NN)
correlations on important quantities,
like the equation of state, density and spin responses or neutrino
opacity, has been firmly recognized\cite{sawyer75,sawyer89,iwamoto82,reddy99,
fantoni01b}. Mean field models where neutron matter is treated
as a weakly interacting system
are insufficient to describe these correlations.

Dense and cold nuclear matter cannot be reproduced in our terrestrial
laboratories. Therefore, it must be studied theoretically, and one needs
to make use of the full machinery of modern many--body theories.

Exotic nuclei with large neutron excess, far from the 
stability valley, including isotopes of nitrogen, oxygen and fluorine, have
been produced in heavy ion reactions with radioactive beams, and 
subsequently studied
spectroscopically. Experimental data on their
binding energies and neutron removal energies provide a very important
body of information on the structure of 
neutron rich matter and on the NN interaction, at least in the 
low density regime. 
It is well known that tensor, spin--orbit and three--body
forces play a fundamental role. Small modifications of these forces may
lead to large effects on the properties of dense and cold hadronic matter.

Neutron droplets were originally introduced
as a homework problem to compare {\sl ab initio} calculations 
with energy density functionals commonly used in astrophysical 
investigations. They also provide clean benchmarks for modern many--body
theories, being simpler systems than either nuclei or nuclear matter.

Closed shell drops made of eight neutrons, 
and open shell ones with seven or six neutrons have been 
studied in the last decade, by using quantum Monte Carlo methods, like
Variational Monte Carlo and Green's function Monte Carlo (GFMC)\cite{smerzi97}.
These quantum simulations carried out with a realistic bare NN interactions 
deviate significantly from energy density functional results\cite{pudliner96}.

Neutron droplet models can also serve to study neutron rich nuclei, devising
one--body effective potentials to describe the interaction
of the halo neutrons with an inert core. For instance, the 
oxygen isotopes can be modeled by neutron droplets 
with up to 10 neutrons out of the $^{16}$O core, and
fluorine isotopes with droplets made of 
one proton and up to 12 neutrons in the halo\cite{pieperPC}.

The recent developments made in Quantum Monte Carlo 
methods, allow for {\sl ab initio} calculations of medium--heavy 
nucleonic systems, at an unprecedented accuracy.
GFMC calculations of the binding energy of nucleonic systems
with $A=14$ have been recently been performed\cite{carlson03}. 
Auxiliary Field Diffusion Monte Carlo (AFDMC) requires order $N^3$
operations per time step and this polynomial scaling allows the
study of many more nucleons. For example,
neutron matter calculations\cite{sarsa03} have used 
114 neutrons in a periodic box to calculate the
equation of state.
In both methods the
spin--isospin dependence of the nucleonic interaction has been handled
without any approximation.

In particular, the AFDMC method looks very promising to perform quantum
simulations of large nuclei and nuclear matter\cite{schmidt99,sarsa03}.
The main reason is because
it samples, rather than performing a complete sum of
the spin states. The sampling is not done directly on the spin states, but
auxiliary fields of the Hubbard--Stratonovich type are introduced to
linearize the spin dependence of the NN potential operator. 
The method consists of a
Monte Carlo sampling of the auxiliary fields, and then propagating 
the spin variables at the sampled values by means
of a rotation of each particle's spin spinor.

In this paper we apply the AFDMC
method to study neutron drops of 7,8 neutrons, kept bound by an effective
one--body potential of the type given in ref.\cite{pudliner96}. 
The main goal is to test the newly developed AFDMC versus other quantum 
Monte Carlo methods, like GFMC or Cluster variational Monte Carlo.
Reasonably good agreement is obtained for the binding energy of $^8n(0^+)$ and 
$^7n(3/2^-)$, whereas  a larger discrepancy exists on the 
spin--orbit splitting of $^7n$.
 
Our calculations represent a preliminary
step towards the study of neutron rich nuclei.

The plan of the paper is the following. In the next section we describe the
hamiltonian used in our calculations.  A brief outline
of the AFDMC method is given in Section III. A reformulation of  
the three--body potentials of the Illinois type is given 
which allows for efficient numerical calculation. 
Results are presented and
discussed in Section IV. The last Section is devoted to conclusions and
future perspectives.  

\section{Hamiltonian}

The ground state properties of the neutron droplets are computed starting from
a nonrelativistic Hamiltonian of the following form:
\begin{eqnarray}
\hat{H} &=& T+V_1+V_2+V_3\nonumber\\\\
&=&-\sum_i\frac{\hbar^2}{2m}\nabla^2_i+\sum_i V_{ext}({\vec r}_i)+
\sum_{i<j} v_{ij} + \sum_{i<j<k}V_{ijk}\nonumber
\end{eqnarray}
The one-body potential $V_{ext}$ stabilizes the drop, 
otherwise unbound. We take it of the same form as in ref. \cite{pudliner96}.
It consists of a Wood-Saxon well of the form
\begin{equation}
 V_{ext}(r) = -\frac{V_0}{1+e^{(r-R)/a}}
\label{extpot}
\end{equation}
where $R=3$fm, and $a=0.65$fm. The parameter $V_0$ determines 
the depth of the well. We have made calculations for 
four different values of $V_0=20,25,30$ and $35$ MeV, where 
the first value has been used in the
GFMC and variational Monte Carlo calculations of ref.\cite{pudliner96,smerzi97} and more recently
in ref.\cite{pieper01}. The larger values of $V_0$ have been considered
to test the reliability of AFDMC for more bound systems, with a large peak
density. To increase the density further $R$ must be decreased.
 
The two--body NN interaction considered 
belongs to the Urbana-Argonne $v_{l}$ potentials :
\begin{equation}
v_l=\sum_{i<j} \sum_{p=1}^l v_p(r_{ij})O^{(p)}(i,j),
\label{pot}
\end{equation}
truncated to include only the following  8 operators ($v'_8$)\cite{argonnev18,
pudliner97,sarsa03}:
\begin{equation}
O^{p=1,8}(i,j)=(1,\vec\sigma_i\cdot\vec\sigma_j,S_{ij},
{\vec L}_{ij}\cdot{\vec S}_{ij})\otimes(1,\vec\tau_i\cdot\vec\tau_j)
\end {equation}
where the operator $S_{ij}=3\vec{\sigma}_i\cdot\hat{r}_{ij}
\vec{\sigma}_j\cdot\hat{r}_{ij}-\vec{\sigma}_i\cdot\vec{\sigma}_j$ is the 
tensor operator 
and $\vec L_{ij}=-\imath\hbar\vec r_{ij}\times (\vec\nabla_i-\vec\nabla_j)/2$
and $\vec S_{ij}=\hbar(\vec\sigma_i+\vec\sigma_j)/2$ are the relative
angular momentum and the total spin for the pair $ij$.
For neutrons $\vec\tau_i\cdot\vec\tau_j=1$, and we are left with an isoscalar 
potential. 

The $v'_8$ potential is a simplified version of the $v_{18}$ potential, having
the same isoscalar parts of $v_{18}$ in all $S$ and $P$ waves, as well as in 
the $^3D_1$ channel and its coupling to the $^3S_1$. It is only semirealistic
because it does not fit
the Nijmegen N--N data \cite{stoks93} 
at a confidence level of $\chi^2/N_{data}\sim 1$, as $v_{18}$ does.
However, the difference between $v_{18}$ and $v'_8$ is rather small for 
densities smaller or of the order of the nuclear matter equilibrium density
$\rho_0=0.16 $fm$^{-3}$, and it can be safely added perturbatively.

The $v'_8$ potential should be considered as a 
{\sl realistic} homework potential, and
it has been used in a number of calculations on light 
nuclei\cite{pudliner97,pieper01}, symmetric nuclear matter\cite{akpa97},
neutron matter\cite{akpa97,sarsa03,baldo04} and spin polarized 
neutron matter\cite{fantoni01}. 

In order to estimate the contribution of the spin--orbit force, we have
also made calculations with a two--body potential obtained from $v'_8$
dropping the spin--orbit terms. We denote this potential as $v'_6$, but
it should not be confused with the $v'_6$ potential given in 
ref.\cite{wipi03} which gives a 
correct binding energy of the deuteron.

The three--body interaction considered encompasses the form of both the
Urbana and Illinois 3-body potentials\cite{pieper01}.
Results will be given for the
Urbana IX potential. The $v'_8$ or the $v'_6$ two--body potentials plus the 
Urbana IX interaction are denoted as $AU8'$ and
$AU6'$ respectively.

\section{Auxiliary Field Diffusion Monte Carlo method}

The Auxiliary Field Diffusion Monte Carlo \cite{schmidt99}, is an extension
to DMC to deal with spin dependent hamiltonians.  
The quadratic dependence of these hamiltonians on the spin operators is
taken care of by sampling  
auxiliary variables, which serve to linearize such dependence through
Hubbard--Stratonovich transformations. A detailed discussion of the method 
can be found in Ref. \cite{fantoni00,sarsa03}. Here we limit ourselves to
briefly outlining the method.
 
The $v'_6$ two--body potential can 
be separated into a spin--independent and a spin--dependent part:
\begin{eqnarray}
V&=&V^{SI}+V^{SD} \ ,\nonumber \\  \nonumber
V^{SD}&=&\sum_{i,j}\sigma_{i\alpha}A_{i\alpha;j\beta}\sigma_{j\beta} \ ,
\end{eqnarray}
where the elements of the matrix $A$ are given by the proper 
combinations of the components $v_p$ in Eq. (\ref{pot}).
Latin indices, like $i$ and $j$, are used to for particles, 
while the greek ones, like $\alpha$ and $\beta$, 
refer to the Cartesian components of the operators. We
use the summation convention  that all repeated greek indices are summed
from 1 to 3.
 
Because $A_{i\alpha;i\beta} =0$ the $3N$ by $3N$ matrix $A$ 
has real eigenvalues and eigenvectors, defined by:
\begin{equation}
\sum_{j}A_{i,\alpha;j\beta}\psi_n^{j\beta}=
\lambda_n\psi_n^{i\alpha} \ .
\end{equation}
The spin--dependent potential can therefore be 
written in terms of such eigenvalues 
and eigenvectors in the following form:
\begin{equation}
V^{SD}=\frac{1}{2}\sum_n\left[\sum_{i,j}\sigma_{i\alpha}\psi_n^
{i\alpha}\lambda_n\psi_n^{j\beta}\sigma_{j\beta}\right] \ .
\end{equation}
If one defines new N-body spin operators 
\begin{equation}
O_n = \sum_i\sigma_{i\alpha}\psi_n^{i\alpha} \ ,
\end{equation}
the spin--dependent  potential becomes
\begin{equation}
V^{SD}=\frac{1}{2}\sum_{n=1}^{3N}\lambda_n O_n^2 \ . 
\end{equation}
In the short--time limit we can decompose  
the imaginary time propagator of the diffusion process, which projects
the ground state out of a trial wavefunction in the following way:
\begin{equation}
e^{-H \Delta \tau} \sim e^{-T \Delta \tau}
e^{-V_c \Delta \tau }e^{-V^{SD} \Delta \tau } \ ,
\label{propagator}
\end{equation}
 
where $V_c=\sum V_{ext}(r_i)+V^{SI}$ is the spin independent part of the
interaction.
The propagation accounting for the kinetic and $V_c$ operators 
gives rise to the usual 
drift--diffusion scheme of DMC. The spin--dependent two--body potential 
part $e^{-V^{SD}\Delta\tau}$ is handled  
by making use of the following Hubbard--Stratonovich transformation
\begin{eqnarray}
& & e^{\displaystyle-\frac{1}{2}\lambda_n O_n^2\Delta \tau} = \nonumber \\
& & \frac{1}{\sqrt{2\pi}}\displaystyle\int_{-\infty}^{+\infty}dx_n\;
e^{\displaystyle -\frac{x_n^2}{2}
-\sqrt{-\lambda_n\Delta\tau} x_n O_n} \ ,
\label{HS}
\end{eqnarray}

with 

\begin{eqnarray}
e^{\displaystyle -V^{SD}\Delta\tau}\sim\prod_n
e^{\displaystyle-\frac{1}{2}\lambda_n O_n^2\Delta \tau} \ ,
\end{eqnarray}

where the commutators amongst the $O_n$ are neglected, which requires to
keep the time step $\Delta\tau$ small enough.

In Eq.(\ref{HS}) the quadratic
dependence on the spin operators is transformed into a linear expression
which corresponds to a rotation in the spin space.
For each eigenvalue $\lambda_n$ a value of $x_n$ is sampled, and
the current spinor value for each particle is multiplied by the set of matrices
given by the transformation in  Eq.(\ref{HS}).
 
The spin--orbit and three-body potentials can be treated within 
the same scheme. It is
important to notice that while the spin--orbit potential is already linear in the
spin operator, it is necessary to eliminate spurious terms from the simple
linearization of the propagator in order to take into account 
corrections at order $\Delta\tau$. This leads to additional 
two-- plus three-body counter terms, which can be treated as additional 
interaction terms\cite{sarsa03}.

\subsection{Three--body potential}

We give in this section a reformulation of 
the Urbana and Illinois 3-body potentials\cite{pieper01} for neutrons,
which provides an efficient and simple way to program them. They  
can be written so that the $O(N^3)$ parts are done with matrix multiplies.
All of these potentials can be written in the form
\begin{eqnarray}
V_{ijk} &=& A_{2\pi}^{PW} O^{2\pi,PW}_{ijk}
+ A_{2\pi}^{SW} O^{2\pi,SW}_{ijk} \nonumber \\
&+& A_{3\pi}^{\Delta R} O^{3\pi,\Delta R}_{ijk}
+A_R O^R_{ijk} \,.
\end{eqnarray}

The 1-pion exchange amplitude in Argonne $v_{18}$ is
\begin{equation}
v_{ij}^\pi = \frac{1}{3} \frac{f_{\pi N N}^2}{4\pi} m_\pi
\vec \tau_i \cdot \vec \tau_j X^{\rm op}_{ij} \ ,
\end{equation}
where
\begin{eqnarray}
X^{\rm op}_{ij} &=& T(m_\pi r_{ij}) \left[ 3
\vec \sigma_i \cdot \hat r_{ij} \vec \sigma_j \cdot \hat r_{ij}
-\vec \sigma_i \cdot \vec \sigma_j \right] \nonumber \\
&+& Y(m_\pi r_{ij}) \vec \sigma_i \cdot \vec \sigma_j \,,
\end{eqnarray}
and the functions $Y(x)$ and $T(x)$ are given by
\begin{eqnarray}
Y(x) &=& \frac{e^{-x}}{x} \xi_Y(r)
\nonumber\\
T(x) &=& \left( \frac{3}{x^2}+\frac{3}{x}+1 \right) Y(x)\xi_T(r)
\nonumber\\
\xi_Y(r) &=& \xi_T(r) = 1-e^{-cr^2} \,.
\end{eqnarray}

The operator $\sum_{i,j} X^{\rm op}_{ij}$ has 
the same algebraic structure as the 
spin dependent part of the two--body potential $V^{SD}$. Therefore, it can
be expressed in a similar way, in terms of a $3N$ by $3N$ matrix
$X_{i\alpha ;j\beta}$ as
\begin{equation}
X^{\rm op}_{ij} = \sigma_{i\alpha} X_{i\alpha ; j\beta} \sigma_{j\beta}\ ,
\end{equation}
with $X_{i\alpha;i\beta} = 0$.  Notice that this matrix
is symmetric under cartesian component interchange 
$\alpha \leftrightarrow \beta$
and under particle label interchange $i \leftrightarrow j$
and is also fully symmetric $X_{j\beta;i\alpha} = X_{i\alpha; j\beta}$.

It is convenient to calculate the square of the $X$ matrix
\begin{equation}
X^2_{i\alpha ; j\beta } = 
\sum_{k} X_{i\alpha;k\gamma} X_{k\gamma;j\beta} \,.
\end{equation}

The Fujita-Miyazawa form for $O^{2\pi,PW}_{ijk}$ and neutrons becomes
\begin{equation}
\sum_{i<j<k} O^{2\pi,PW}_{ijk} =
4 \sum_{i<j} \sigma_{i\alpha}\sigma_{j\beta} X^2_{i\alpha;j\beta}\ .
\end{equation}

The spin independent $O^{R}_{ijk}$ terms can be written 
using just pair sums as
\begin{eqnarray}
G^R_i &=&\sum_{k \neq i} T^2(m_\pi r_{ik})
\nonumber\\
G^R_0 &=&-\sum_{i<j} T^4(m_\pi r_{ij})
\nonumber\\
\sum_{i<j<k} O^R_{ijk}&=& G^R_0+\frac{1}{2} \sum_i [G^R_i]^2\ .
\end{eqnarray}

The Tucson S wave component can be written in terms of
\begin{eqnarray}
Z(x) &=& \frac{x}{3} \left [Y(x)-T(x) \right ]
\nonumber\\
\vec G^S_{i j} &=& \vec r_{ij} Z(m_\pi r_{ij})
\nonumber\\
\vec G^S_{i i} &=& 0 \,.
\end{eqnarray}
The two--pion, S wave component of the potential becomes
\begin{equation}
\sum_{i<j<k} O^{2\pi,SW}_{ijk} = \sum_{i<j} \sigma_{i\alpha}\sigma_{j\beta}
\sum_k G^S_{\alpha ;ik} G^S_{\beta ;jk} \,.
\end{equation}
Note the inner sum makes this a matrix multiply. This and the $X^2$
calculation above are the only order $N^3$ operations needed.

The 3-pion terms have a central part and a spin dependent part.
\begin{equation}
\sum_{i<j<k} O^{3\pi,\Delta R}_{ijk} = V_c + V_s\ .
\end{equation}
The central part is
\begin{equation}
V_c = \frac{400}{9}\sum_{i<j} X^2_{i\alpha;j\beta}X_{i\alpha;j\beta}\ ,
\end{equation}
which is $200/9$ times the trace of $X^3$.
The spin dependent part is
\begin{equation}
V_s = \frac{200}{27}
\sum_{i<j} \sigma_{i\alpha} \sigma_{j\beta} X^2_{i\gamma;j\kappa}
X_{i\delta;j\omega}
\epsilon_{\alpha\gamma\delta} \epsilon_{\beta \kappa \omega}\  ,
\end{equation}
and one can just write out the 4 nonzero terms for each combination
of $\sigma_i$ and $\sigma_j$.

As might be expected, the expression used above is related to $S^D_{ijk}$
of ref.\cite{pieper01} if only the $k$ term of the $X^2$ sum is kept,
\begin{equation}
3 S^D_{jki}=
\sigma_{i\alpha} \sigma_{j\beta} X_{i\gamma;k\phi} X_{k\phi;j\kappa}
X_{i\delta;j\omega}
\epsilon_{\alpha\gamma\delta} \epsilon_{\beta \kappa \omega} \,.
\end{equation}
The $I$ part is
\begin{equation}
3S^I_\sigma = 3\sum_{\rm cyc} S^{D}_{ijk} +
2 X_{i\alpha; j\beta} X_{i\beta; j\gamma} X_{i\gamma; j\alpha} \,.
\end{equation}
Inserting 1 or 3 Levi-Civita symbols instead of 2
will produce the $A$ operators of ref.\cite{pieper01},
\begin{eqnarray}
3A_\sigma^I &=& \imath \sigma_{i\alpha} \sigma_{j \beta} \sigma_{k\mu}
X_{i\gamma; j\kappa} X_{j\omega; k\nu} X_{i\rho; k\delta} \nonumber \\
&\times& \epsilon_{\alpha \gamma \delta} \epsilon_{\beta \kappa \omega}
\epsilon_{\mu \nu \rho}
\nonumber\\
3A^D_{\sigma,ijk} &=& -i \sigma_{1\alpha} 
X_{i\gamma; j\beta} X_{j\beta; k\mu} X_{i\mu; k\delta}
\epsilon_{\alpha \gamma \delta } \,.
\end{eqnarray}

The spin dependent part of the three--body interaction can be easily
included in the matrix $A_{i\alpha;j\beta}$ by
\begin{eqnarray}
A_{i\alpha;j\beta} &\rightarrow& A_{i\alpha;j\beta}+2 A_{2\pi}^{PW}
X^2_{i\alpha;j\beta} \nonumber \\
&+& \frac{1}{2} A_{2\pi}^{SW} \sum_k G^S_{\alpha;ik} G^S_{\beta;jk}
\nonumber \\
&+&\frac{200}{54} A_{3\pi}^{\Delta R}X^2_{i\gamma;j\kappa} X_{i\delta;j\omega}
\epsilon_{\alpha\gamma\delta} \epsilon_{\beta \kappa \omega}\  .
\end{eqnarray} 

\subsection{Trial wave function}

The wave function used as a trial and importance function for the DMC algorithm
has the following form
\begin{eqnarray}
\Psi(R,S) =  F_J(R)\ D(R,S) \ ,
\end{eqnarray}
where $ R\equiv (\vec r_1,\dots,\vec r_N) $ and $S\equiv (s_1,\dots ,s_N)$.
The spin assignements $s_i$ consist in giving the spinor components, namely
\[
s_i \equiv \left(\begin{array}{c} 
u_i \\ d_i
\end{array}\right)=u_i |\uparrow\rangle + d_i |\downarrow\rangle \ , 
\]
where $u_i$ and $d_i$ are complex numbers. 
The Jastrow correlation operator is given by
\begin{eqnarray}
F_J&=&\prod_{i<j}f_J(r_{ij})\ ,
\end{eqnarray}
and
\begin{eqnarray} 
D(R,S) &=& {\rm det} M
\end{eqnarray}
is the Slater determinant
where the elements of the Slater matrix are
one--body spin--space orbitals,
\begin{equation}
M_{ij} = \langle \vec r_i,s_i | \phi_j\rangle \,.
\end{equation}

The Jastrow function $f_J$ has been taken as the scalar component of the
FHNC/SOC correlation operator $\hat F_{ij}$ which minimizes the energy per
particle of neutron matter at density 
$\rho=0.16fm^{-3}$ \cite{wiringa88,sarsa03}.
The single particle orbitals are taken as
solutions of the Schr\"oedinger equation
\begin{equation}
\left[-\frac{\hbar^2}{2m}\nabla^2+V_{WS}(\vec{r})\right]\phi(\vec{r})=
\epsilon\phi(\vec{r}),
\end{equation}
where $V_{WS}$ is a Wood-Saxon well, like $V_{ext}$ of Eq. (\ref{extpot}) 
with parameters chosen to optimize the expectation value of the energy 
\begin{equation}
E_T=\frac{\langle\Psi|\hat H\Psi\rangle}{\langle \Psi|\Psi\rangle},
\end{equation}
for each value of $V_0$ considered. $E_T$ is calculated with variational
Monte Carlo.
The parameters of the resulting Wood--Saxon potential are reported in
Table \ref{table:param}. 

\begin{table}
\caption{ Variational parameters $V_0^{var}$, $R^{var}$ and $a^{var}$ 
of the Wood--Saxon potential $V_{WS}$ found
in correspondence of the various depths $V_0$ of the well of $V_{ext}$.
}
\begin{center}
\begin{tabular}{lcccc}
\toprule
 $V_0 $(MeV)     & 20 &  25  &  30  &  35  \\
\colrule
$V_0^{var} $(MeV)& 20.0 & 25.0   & 30.0    & 35.0     \\     
$R^{var} $(fm)   & 1.5    & 1.6   & 1.4    & 1.2    \\
$a^{var} $(fm)   &  0.60  & 0.80   & 0.80    & 0.80     \\
\botrule
\end{tabular}
\end{center}
\label{table:param}
\end{table}

In the case of $^8n(0^+)$
the orbitals fill the $s$ and $p$ shells. 
In the case of $^7n(3/2^-)$ the missing orbital off the $p$--shell is 
in the state $|l,m,s=1,1,1/2\rangle$; for  $^7n(1/2^-)$ we have considered
the combination  $(|1,1,-1/2\rangle-|1,0,1/2\rangle)/\sqrt{2}$.  
The above choices give rise to different determinants $D_0$, $D_{3/2}$
and $D_{1/2}$ respectively. They are
evaluated at the current values of the positions and spin assignments
of the nucleons in the walker $|R,S\rangle$.

The Jastrow operator $F_J$ misses all the spin--dependent correlations.
Therefore our trial function is rather poor and we do not expect particularly 
good results for our mixed estimates, except for the total energy. 
Better trial functions 
can be considered, although they require a more demanding computational effort. 
For instance, the space--spin orbitals can be
modified to include spin correlations in the Slater determinant, as
recently done in neutron matter calculation to take into account the
spin--orbit correlations in the trial function\cite{brualla03}. 
Such improved trial
function provides a better nodal surface, and gives a better variance.

\subsection{The algorithm}

The AFDMC algorithm is implemented as usual,
with a propagation in imaginary time of a population of walkers $|R,S\rangle$
according to the propagator in eq. (\ref{propagator})
with the standard drift-diffusion procedure. In addition one has to sample the
$x_n$ auxiliary variables given in Eq.(\ref{HS}) to rotate the spinors. 
After that all the weight factors are computed, 
they are combined to evaluate a new value of $\langle\Psi|R,S\rangle$.
 
In order to avoid the fermion sign problem due to 
the antisymmetric character of the
wave function, a path constraint is introduced. If the real part of 
$\langle\Psi|R,S\rangle$ is negative, the walker is included in the evaluation
of the mixed and growth energies, but then is dropped from the population. 
In general,
the importance sampling makes the number of dropped walkers small.
In our calculations here the number of node crossings is of
order 1 percent. 

\section{Results}

In Table \ref{table:ener} we show our AFDMC results  for the mixed energies, 
obtained for a $^8n(0^+)$ droplet
confined by a Wood--Saxon well $V_{ext}$ with a depth value 
$V_0=20 $MeV in correspondence with the AU8' and AU6' nuclear interactions.
They are compared with the most recent GFMC estimates\cite{pieper01}\footnote{
The GFMC results for the AU8' interaction are not given explicitely in 
ref.\cite{pieper01}. They are extracted 
by adding to AU18 energies the energy differences relative to Argonne $v'_8$ and
Argonne $v_{18}$ with no three--body force.} 

One can see that there is an overall agremeent between AFDMC and GFMC.
The discrepancies on the binding energies of $^8n (0^+)$ and
$^7n (3/2^-)$ are within $2\%$. This is very small if one consider that the 
binding energy results from a large cancellation between potential and kinetic 
energy (the mixed value of the potential energy is of $-124.6(3) $MeV) and that
the variational energy with our $\Psi$ is only $ -26.27 $MeV .

A larger discrepancy is obtained for the 
$^7n (1/2^-)$ state, which makes the AFDMC spin--orbit splitting of $^7n$ about half that of GFMC. 
Such discrepancy may be due to differences between AFDMC and GFMC in treating the
spin--orbit interaction, as already pointed out by the neutron matter calculations
of ref.\cite{sarsa03}.

We have attempted to go beyond our path constraint, 
by doing transient estimation.
This requires the guiding function to have no zeroes where the true ground 
state is non zero. The guiding function that we have used is the modulus 
of the antisymmetric function smoothed with a term that decays to zero as
function of the distance of neutrons from the center of the drop: 

\begin{equation}
\Psi_G = \sqrt{\Psi^2 +\alpha^2\nu(\vec r_1\dots \vec r_N)}\ ,
\end{equation}

where:

\begin{equation}
\nu(\vec r_1\dots \vec r_N) = \prod_{i=1}^N\left(\sum_{\sigma}\sum_{k=1}^N|
\langle \vec r_i \sigma |\phi_k\rangle|^2\right)\ . 
\end{equation}

The parameter $\alpha$ has been chosen equal to 0.1 .
Branching is done on the magnitude of the total weight 
$w_k$ of the k--th walker.  The mixed estimate is given by
\begin{equation}
O_{\rm mixed} = \frac{
\sum_k w_k
\frac{\langle \Psi |O|R_kS_k\rangle}{\langle \Psi |R_kS_k\rangle}
\frac{\langle \Psi |R_kS_k\rangle}{\langle \Psi_G |R_kS_k\rangle}
}{
\sum_k w_k
\frac{\langle \Psi |R_kS_k\rangle}{\langle \Psi_G |R_kS_k\rangle}
}\ .
\end{equation}

The results obtained for the $^8n(0^+)$ drop are shown
in Fig. 1. As it can be seen, the energy in the imaginary time interval
of 0.04 MeV$^{-1}$ does not show any significant decay, and it is compatible
within errorbars with the estimate of the constrained path energy which is
$-37.8\  $MeV. 
\begin{figure}
\includegraphics[angle=270,scale=0.32]{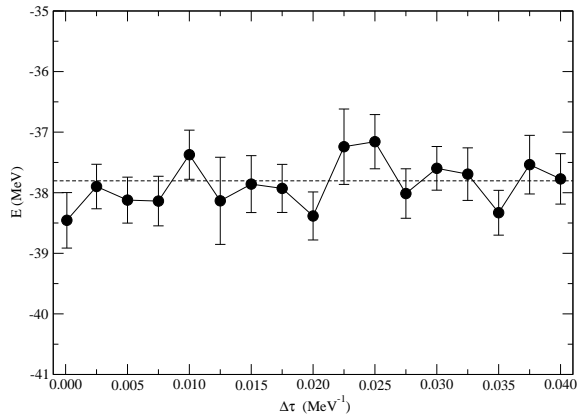}
\caption{Results of a transient estimation for the ground state energy of
a $^8$n started from an AFDMC calculation.}
\end{figure}
The errorbars are in any case rather large, which is 
due to a poor description of
spin dependent correlations in both trial and guiding functions.
They can be included by using the backflow
form of ref.\cite{brualla03}. Work in this direction is in progress. 

Another interesting comparison is with the results obtained 
with various energy density functional models, like (i) Skyrme M\cite{krivine80}, 
(ii) Skyrme 1'\cite{vautherin72,ravenhall72}, (iii) FPS\cite{pandharipande89} and 
(iv) FPS--21\cite{pethick95}, which reproduce the ground state energies of stable
closed shell nuclei rather accurately. 
The energy density functional models provide a range of values for the binding energy of $^8n(0^+)$ which
goes from $-32.1 $MeV of FPS to $-47.4 $MeV of Skyrme M\cite{pudliner96}. Moreover,
the $^7n$ spin--orbit splitting is about $3 $MeV, too large with respect to the quantum Monte Carlo
estimates.   

\begin{table}
\caption{Ground state AFDMC energies of $^8n(0^+)$, $^7n(\frac{1}{2}^-)$
and $^7n(\frac{3}{2}^-)$ droplets for 
$V_0=20 $MeV and the AU8' and AU6' interactions. The cluster
variational Monte Carlo (CVMC) and GFMC results of
ref.\cite{pieper01} for the AU8' and the full AU18 (Argonne $v_{18}$ plus Urbana IX)
are also reported for comparison. The last column reports the spin--orbit
splittings (SOS) in MeV of $^7n$, given by the energy difference 
between the $^7n(\frac{3}{2}^-)$ and $^7n(\frac{1}{2}^-)$ states. 
}
\begin{center}
\begin{tabular}{lcccc}
\toprule
              & $^8n(0^+)$ & $^7n(\frac{1}{2}^-)$ & $^7n(\frac{3}{2}^-)$& SOS \\
\colrule
GFMC(AU18)    & -37.8(1)   &  -33.2(1)     &  -31.7(1) & 1.5(2)  \\     
CVMC(AU18)    & -35.5(1)   &  -31.2(1)     &  -29.7(1) & 1.5(2)  \\
GFMC(AU8')    & -38.3(1)   &  -34.0(1)     &  -32.4(1) & 1.6(2)  \\
AFDMC(AU8')   & -37.8(2)   &  -32.5(2)     &  -31.8(1) & 0.7(2)  \\
AFDMC(AU6')   & -36.82(5)  &  -31.6(2)     &  -30.8(4) & 0.8(5)  \\
\botrule
\end{tabular}
\end{center}
\label{table:ener}
\end{table}

We have repeated the calculations of  ground state energies and densities in the 
Skyrme 1' and in the SKM models for the $^8n(0^+)$ drop. 
The interaction parameters have been taken from Ref.\cite{pethick95}.  
The energies found are $-47.03$MeV and $-50.6$MeV respectively.
In Fig. 2 we compare the neutron density profiles 
of the $^8n(0^+)$ drop obtained from AFDMC
calculations with $AU8'$ and $AU6'$ interactions, and from self consistent mean field
calculations using the Skyrme 1' model, and the SKM model. 

As it can be seen, both of the energy density functional results 
differ significantly from the  quantum Monte Carlo ones, particularly
in the peak density value, at the center of the drop.
The SKM however shows a density which is much closer to 
the quantum Monte Carlo result than the Skyrme 1' one.

\begin{figure}
\includegraphics[angle=270,scale=0.32]{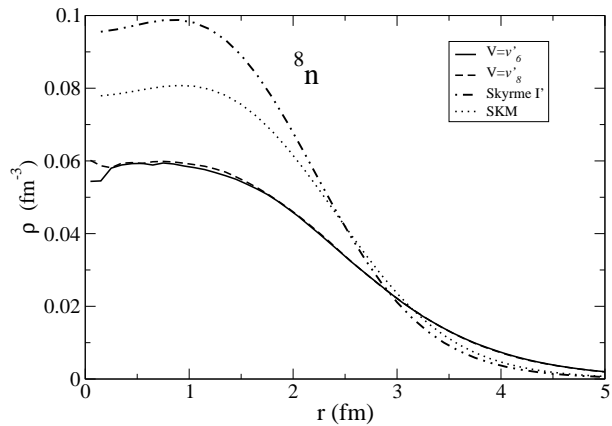}
\caption{Density profiles for a $^8n$ drop for $V_0=20$MeV. Solid line:
result with the $v^{8'}$ interaction; dashed line: result with the 
$v^{6'}$ interaction; dotted-dashed line: HF with Skyrme 1'; dotted line: SKM. 
 }
\end{figure}

In Table \ref{table:display}  we report the  results of 
AFDMC calculations for a $^8$n droplet for
different values of the depth of the confining potential $V_0$. The importance function
for each case was optimized only with respect to the parameters of the potential well
from which  the orbitals to be included in the Slater determinant are obtained.  
\begin{table}
\caption{AFDMC ground state energy (in MeV) for a $^8$n droplet as a function of
the parameter $V_0$ (in MeV) of $V_{ext}$. 
The energy difference $\Delta E=E(AU8')-E(AU6')$ gives an 
unbiased estimate of the spin--orbit interaction contribution.}
\begin{center}
\begin{tabular}{cccccccc}
\toprule
$V_0$ & $E_T$& $E_{AU8'}$ & $\langle V \rangle$ & $\langle 
v_{\vec L\cdot\vec S}\rangle$ &$E_{AU6'}$&$\Delta E$\\
\colrule
20&-26.27(7)&-37.8(2)&-124.6(3)&-0.41(7)&-36.82(5)&-1.0(2)\\
25&-48.0(3)&-60.5(1)&-168.2(4)&-0.59(4)&-59.6(1)&-0.9(2)\\
30&-69.3(3)&-85.3(2)&-201.3(2)&-0.72(4)&-83.7(1)&-1.6(4)\\
35&-92.7(3)&-111.0(2)&-232.4(2)&-0.71(3)&-109.86(5)&-1.1(4)&\\
\botrule
\end{tabular}
\end{center}
\label{table:display}
\end{table}

Together with the AFDMC energies
we report the variational energies $E_T$. 
The difference between tha variational and the AFDMC energy is
quite large, in all cases between 20 and 30\%. This 
indicates that the evaluation of mixed estimators 
is at the same level of accuracy for
all the cases, and the observed trends should not be 
too affected from the particular 
choice of the trial functions.
The mixed energies are given for the $AU8'$ and  
$AU6'$ interactions,  
in order to calculate the contributions of the spin orbit interaction
to the total energies. These are given in by $\Delta E$ on the last 
column of the figure.

We have also evaluated the mixed estimate of the spin--orbit interaction.
Since $\langle\Psi|v_{\vec L\cdot\vec S} |\Psi\rangle=0$, the spin--orbit
expectation value $ \langle v_{\vec L\cdot\vec S}\rangle$ is approximately
given by half the mixed estimate, and is reported in the 
fifth column of the Table.
It is not very sensible to 
the depth of the potential well, in contrast with the total 
energy, which varies almost linearly with it.

One can see that $ \langle v_{\vec L\cdot\vec S}\rangle$ is about
60\% of $\Delta E$ in the range of depths of the well considered.
On the other hand this value is very close to the spin--orbit splitting evaluated in the
$^7n$ drop reported in Table II. It should also be noticed that in the latter 
case, the values of the spin--orbit splitting obtained with the AU8' and the AU6' are
very similar. This means that the evaluation of the spin--orbit splitting in $^7n$
is not strongly influenced by the spin--orbit term in the interaction. 
The spin--orbit splitting found in our AFDMC is smaller (about one half) than the values reported
from GFMC calculations. However, the consistencies in our results
mentioned above makes us confident in the robustness of the
evaluation of spin--orbit contributions in our method.

\begin{figure}
\includegraphics[angle=270,scale=0.28]{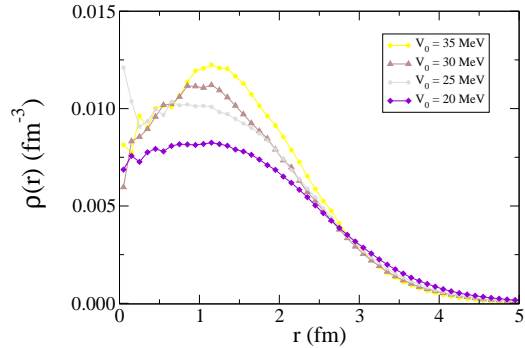}
\caption{Neutron density profiles in a $^8n$ drop for different values of the
depth of the confining potential $V_0$}
\end{figure}

In order to understand the connection between the 
parameter $V_0$ and the neutron
density inside the drop, we have plotted 
in Fig. 3 the mixed estimators for the density
profiles for all the confinements considered. 
Although the density at the center remains 
in a range $0.06<\rho(0)<0.09$fm$^{-3}$, it shows  a 
larger dip at the center, and, consequently a 
stronger localization of neutrons
at a distance of $\sim1 $fm, for larger values of $V_0$. 

\section{Conclusions and perspectives}

In this paper we have applied the Auxiliary Field 
Diffusion Monte Carlo method to the
study of a closed-shell neutron drop $^8n(0^+)$ and  of the open shells
$^7n(1/2^-)$  and $^7n(3/2^-)$ confined by an external Wood--Saxon field.
The method compares rather well
with standard GFMC techniques. Since the AFDMC,   
scales better than GFMC with the number of particles it is currently 
the only viable quantum Monte Carlo method to study large neutron rich systems.

We have made a detailed analysis of the
contribution to the total energy of the spin-orbit interaction, which 
previous calculations on neutron matter\cite{sarsa03} has indicated as
the main difference between AFDMC and GFMC calculations.
Such analysis has been carried on
at various values of the depth $V_0$ of the confining
potential $V_{ext}$, and, therefore, for different confined neutron 
structures. We have found a
spin-orbit contribution which is generally smaller than in 
GFMC calculations, confirming the findings of ref.\cite{sarsa03}. 
Such contribution lowers the 
binding energy of about 1 MeV in all considered cases.

The results obtained for the binding energies of the $^7n(1/2^-)$  
and $^7n(3/2^-)$ droplets confirm earlier GFMC results on the
spin orbit splittings, which are significantly smaller than those
provided by the energy density functional models.
The AFDMC value of the spin--orbit splitting is about one half of the value predicted by GFMC.

Future work includes improving the trial function $\Psi$, as well as
the guiding function $\Psi_G$ for transient estimate.
The properties of oxygen and flourine istopes can be calculated with
the AFDMC method.

\begin{acknowledgments}

S. F. wish to thank the International Centre for Theoretical Physics in 
Trieste for partial support. 
A. S. acknowledges the Spanish Ministerio 
de Ciencia y Tecnologia for partial support under contract BMF2002-00200
Calculations have been performed in part on the ALPS cluster at ECT* in Trento.

\end{acknowledgments}

\bibliography{neutron}

\begin{thebibliography}{10}
\expandafter\ifx\csname bibnamefont\endcsname\relax
  \def\bibnamefont#1{#1}\fi
\expandafter\ifx\csname bibfnamefont\endcsname\relax
  \def\bibfnamefont#1{#1}\fi
\expandafter\ifx\csname url\endcsname\relax
  \def\url#1{\texttt{#1}}\fi
\expandafter\ifx\csname urlprefix\endcsname\relax\def\urlprefix{URL }\fi
\expandafter\ifx\csname bibinfo\endcsname\relax \def\bibinfo#1#2{#2}\fi
\expandafter\ifx\csname eprint\endcsname\relax \def\eprint#1{#1}\fi

\bibitem{pethick95}
\bibinfo{author}{\bibfnamefont{C.~J.} \bibnamefont{Pethick}},
  \bibinfo{author}{\bibfnamefont{D.~G.} \bibnamefont{Ravenhall}},
  \bibnamefont{and} \bibinfo{author}{\bibfnamefont{C.~P.}
  \bibnamefont{Lorenz}}, \bibinfo{journal}{Nucl. Phys. A}
  \textbf{\bibinfo{volume}{584}}, \bibinfo{pages}{675} (\bibinfo{year}{1982}).

\bibitem{raffelt96}
\bibinfo{author}{\bibfnamefont{G.~G.} \bibnamefont{Raffelt}},
  \emph{\bibinfo{title}{The stars as laboratories of fundamental physics}}
  (\bibinfo{publisher}{University of Chicago}, \bibinfo{address}{Chicago \&
  London}, \bibinfo{year}{1996}).

\bibitem{kratz93}
\bibinfo{author}{\bibfnamefont{K.~L.} \bibnamefont{Kratz}},
  \bibinfo{author}{\bibfnamefont{J.~P.} \bibnamefont{Bitouzet}},
  \bibinfo{author}{\bibfnamefont{F.~K.} \bibnamefont{Thielemann}},
  \bibinfo{author}{\bibfnamefont{P.}~\bibnamefont{Moeller}}, \bibnamefont{and}
  \bibinfo{author}{\bibfnamefont{B.}~\bibnamefont{Pfeiffer}},
  \bibinfo{journal}{Astrophys. J.} \textbf{\bibinfo{volume}{403}},
  \bibinfo{pages}{216} (\bibinfo{year}{1993}).

\bibitem{akmal98}
\bibinfo{author}{\bibfnamefont{A.}~\bibnamefont{Akmal}},
  \bibinfo{author}{\bibfnamefont{V.~R.} \bibnamefont{Pandharipande}},
  \bibnamefont{and} \bibinfo{author}{\bibfnamefont{D.~G.}
  \bibnamefont{Ravenhall}}, \bibinfo{journal}{Phys. Rev. C}
  \textbf{\bibinfo{volume}{58}}, \bibinfo{pages}{1804} (\bibinfo{year}{1998}).

\bibitem{lattimer01}
\bibinfo{author}{\bibfnamefont{J.~M.} \bibnamefont{Lattimer}} \bibnamefont{and}
  \bibinfo{author}{\bibfnamefont{M.}~\bibnamefont{Prakash}},
  \bibinfo{journal}{Astrophys. J.} \textbf{\bibinfo{volume}{550}},
  \bibinfo{pages}{426} (\bibinfo{year}{2001}).

\bibitem{sawyer75}
\bibinfo{author}{\bibfnamefont{R.~F.} \bibnamefont{Sawyer}},
  \bibinfo{journal}{Phys. Rev. D} \textbf{\bibinfo{volume}{11}},
  \bibinfo{pages}{2740} (\bibinfo{year}{1975}).

\bibitem{sawyer89}
\bibinfo{author}{\bibfnamefont{R.~F.} \bibnamefont{Sawyer}},
  \bibinfo{journal}{Phys. Rev. C} \textbf{\bibinfo{volume}{40}},
  \bibinfo{pages}{865} (\bibinfo{year}{1989}).

\bibitem{iwamoto82}
\bibinfo{author}{\bibfnamefont{N.}~\bibnamefont{Iwamoto}} \bibnamefont{and}
  \bibinfo{author}{\bibfnamefont{C.~J.} \bibnamefont{Pethick}},
  \bibinfo{journal}{Phys. Rev. D} \textbf{\bibinfo{volume}{25}},
  \bibinfo{pages}{313} (\bibinfo{year}{1982}).

\bibitem{reddy99}
\bibinfo{author}{\bibfnamefont{S.}~\bibnamefont{Reddy}},
  \bibinfo{author}{\bibfnamefont{M.}~\bibnamefont{Prakash}},
  \bibinfo{author}{\bibfnamefont{J.~M.} \bibnamefont{Lattimer}},
  \bibnamefont{and} \bibinfo{author}{\bibfnamefont{J.~A.} \bibnamefont{Pons}},
  \bibinfo{journal}{Phys. Rev C} \textbf{\bibinfo{volume}{59}},
  \bibinfo{pages}{2888} (\bibinfo{year}{1999}).

\bibitem{fantoni01b}
\bibinfo{author}{\bibfnamefont{S.}~\bibnamefont{Fantoni}},
  \bibinfo{author}{\bibfnamefont{A.}~\bibnamefont{Sarsa}}, \bibnamefont{and}
  \bibinfo{author}{\bibfnamefont{K.~E.} \bibnamefont{Schmidt}},
  \bibinfo{journal}{Phys. Rev. Lett.} \textbf{\bibinfo{volume}{87}},
  \bibinfo{pages}{181101} (\bibinfo{year}{2001}).

\bibitem{smerzi97}
\bibinfo{author}{\bibfnamefont{A.}~\bibnamefont{Smerzi}},
  \bibinfo{author}{\bibfnamefont{D.~G.} \bibnamefont{Ravenhall}},
  \bibnamefont{and} \bibinfo{author}{\bibfnamefont{V.~R.}
  \bibnamefont{Pandharipande}}, \bibinfo{journal}{Phys. Rev. C}
  \textbf{\bibinfo{volume}{56}}, \bibinfo{pages}{2549} (\bibinfo{year}{1997}).

\bibitem{pudliner96}
\bibinfo{author}{\bibfnamefont{B.~S.} \bibnamefont{Pudliner}},
  \bibinfo{author}{\bibfnamefont{A.}~\bibnamefont{Smerzi}},
  \bibinfo{author}{\bibfnamefont{J.}~\bibnamefont{Carlson}},
  \bibinfo{author}{\bibfnamefont{V.~R.} \bibnamefont{Pandharipande}},
  \bibinfo{author}{\bibfnamefont{S.~C.} \bibnamefont{Pieper}},
  \bibnamefont{and} \bibinfo{author}{\bibfnamefont{D.~G.}
  \bibnamefont{Ravenhall}}, \bibinfo{journal}{Phys. Rev. Lett.}
  \textbf{\bibinfo{volume}{76}}, \bibinfo{pages}{2416} (\bibinfo{year}{1996}).

\bibitem{pieperPC}
\bibinfo{author}{\bibfnamefont{S.~C.} \bibnamefont{Pieper}},
  \bibinfo{journal}{private communication}  (\bibinfo{year}{2003}).

\bibitem{carlson03}
\bibinfo{author}{\bibfnamefont{J.}~\bibnamefont{Carlson}},
  \bibinfo{author}{\bibfnamefont{J.}~\bibnamefont{{Morales, Jr.}}},
  \bibinfo{author}{\bibfnamefont{V.~R.} \bibnamefont{Pandharipande}},
  \bibnamefont{and} \bibinfo{author}{\bibfnamefont{D.~G.}
  \bibnamefont{Ravenhall}}  (\bibinfo{year}{2003}), \eprint{nucl-the/0303041}.

\bibitem{sarsa03}
\bibinfo{author}{\bibfnamefont{A.}~\bibnamefont{Sarsa}},
  \bibinfo{author}{\bibfnamefont{S.}~\bibnamefont{Fantoni}},
  \bibinfo{author}{\bibfnamefont{K.~E.} \bibnamefont{Schmidt}},
  \bibnamefont{and} \bibinfo{author}{\bibfnamefont{F.}~\bibnamefont{Pederiva}}
  (\bibinfo{year}{2003}), \eprint{nucl-th/0303035}.

\bibitem{schmidt99}
\bibinfo{author}{\bibfnamefont{K.~E.} \bibnamefont{Schmidt}} \bibnamefont{and}
  \bibinfo{author}{\bibfnamefont{S.}~\bibnamefont{Fantoni}},
  \bibinfo{journal}{Phys. Lett. B} \textbf{\bibinfo{volume}{446}},
  \bibinfo{pages}{93} (\bibinfo{year}{1999}).

\bibitem{pieper01}
\bibinfo{author}{\bibfnamefont{S.~C.} \bibnamefont{Pieper}},
  \bibinfo{author}{\bibfnamefont{V.~R.} \bibnamefont{Pandharipande}},
  \bibinfo{author}{\bibfnamefont{R.~B.} \bibnamefont{Wiringa}},
  \bibnamefont{and} \bibinfo{author}{\bibfnamefont{J.}~\bibnamefont{Carlson}},
  \bibinfo{journal}{Phys. Rev. C} \textbf{\bibinfo{volume}{64}},
  \bibinfo{pages}{14001} (\bibinfo{year}{2001}).

\bibitem{argonnev18}
\bibinfo{author}{\bibfnamefont{R.~B.} \bibnamefont{Wiringa}},
  \bibinfo{journal}{Argonne $v_{18}$ and $v_8'$ Potential Package,
  http://www.phy.anl.gov/theory/research/ av18/av18pot.f}
  (\bibinfo{year}{1994}).

\bibitem{pudliner97}
\bibinfo{author}{\bibfnamefont{B.~S.} \bibnamefont{Pudliner}},
  \bibinfo{author}{\bibfnamefont{V.~R.} \bibnamefont{Pandharipande}},
  \bibinfo{author}{\bibfnamefont{J.}~\bibnamefont{Carlson}},
  \bibinfo{author}{\bibfnamefont{S.~C.} \bibnamefont{Pieper}},
  \bibnamefont{and} \bibinfo{author}{\bibfnamefont{R.~B.}
  \bibnamefont{Wiringa}}, \bibinfo{journal}{Phys. Rev C}
  \textbf{\bibinfo{volume}{56}}, \bibinfo{pages}{1720} (\bibinfo{year}{1997}).

\bibitem{stoks93}
\bibinfo{author}{\bibfnamefont{V.~G.~J.} \bibnamefont{Stoks}},
  \bibinfo{author}{\bibfnamefont{R.~A.~M.} \bibnamefont{Klomp}},
  \bibinfo{author}{\bibfnamefont{M.~C.~M.} \bibnamefont{Rentmeester}},
  \bibnamefont{and} \bibinfo{author}{\bibfnamefont{J.~J.} \bibnamefont{{de
  Swart}}}, \bibinfo{journal}{Phys. Rev C} \textbf{\bibinfo{volume}{48}},
  \bibinfo{pages}{792} (\bibinfo{year}{1993}).

\bibitem{akpa97}
\bibinfo{author}{\bibfnamefont{A.}~\bibnamefont{Akmal}} \bibnamefont{and}
  \bibinfo{author}{\bibfnamefont{V.~R.} \bibnamefont{Pandharipande}},
  \bibinfo{journal}{Phys. Rev. C} \textbf{\bibinfo{volume}{56}},
  \bibinfo{pages}{2261} (\bibinfo{year}{1997}).

\bibitem{baldo04}
\bibinfo{author}{\bibfnamefont{M.}~\bibnamefont{Baldo}} \bibnamefont{and}
  \bibinfo{author}{\bibfnamefont{C.}~\bibnamefont{Maieron}},
  \bibinfo{journal}{Phys. Rev. C} \textbf{\bibinfo{volume}{69}},
  \bibinfo{pages}{014301} (\bibinfo{year}{2004}).

\bibitem{fantoni01}
\bibinfo{author}{\bibfnamefont{S.}~\bibnamefont{Fantoni}} \bibnamefont{and}
  \bibinfo{author}{\bibfnamefont{K.~E.} \bibnamefont{Schmidt}},
  \bibinfo{journal}{Nucl. Phys. A} \textbf{\bibinfo{volume}{690}},
  \bibinfo{pages}{456} (\bibinfo{year}{2001}).

\bibitem{wipi03}
\bibinfo{author}{\bibfnamefont{R.~B.} \bibnamefont{Wiringa}} \bibnamefont{and}
  \bibinfo{author}{\bibfnamefont{S.~C.} \bibnamefont{Pieper}},
  \bibinfo{journal}{Phys. Rev. Lett.} \textbf{\bibinfo{volume}{89}},
  \bibinfo{pages}{182501} (\bibinfo{year}{2003}).

\bibitem{fantoni00}
\bibinfo{author}{\bibfnamefont{S.}~\bibnamefont{Fantoni}},
  \bibinfo{author}{\bibfnamefont{A.}~\bibnamefont{Sarsa}}, \bibnamefont{and}
  \bibinfo{author}{\bibfnamefont{K.~E.} \bibnamefont{Schmidt}},
  \bibinfo{journal}{Prog. Part. Nucl. Phys.} \textbf{\bibinfo{volume}{44}},
  \bibinfo{pages}{63} (\bibinfo{year}{2000}).

\bibitem{wiringa88}
\bibinfo{author}{\bibfnamefont{R.~B.} \bibnamefont{Wiringa}},
  \bibinfo{author}{\bibfnamefont{V.}~\bibnamefont{Fiks}}, \bibnamefont{and}
  \bibinfo{author}{\bibfnamefont{A.}~\bibnamefont{Fabrocini}},
  \bibinfo{journal}{Phys. Rev. C} \textbf{\bibinfo{volume}{38}},
  \bibinfo{pages}{1010} (\bibinfo{year}{1988}).

\bibitem{brualla03}
\bibinfo{author}{\bibfnamefont{L.}~\bibnamefont{Brualla}},
  \bibinfo{author}{\bibfnamefont{S.~A.} \bibnamefont{Vitiello}},
  \bibinfo{author}{\bibfnamefont{A.}~\bibnamefont{Sarsa}},
  \bibinfo{author}{\bibfnamefont{S.}~\bibnamefont{Fantoni}}, \bibnamefont{and}
  \bibinfo{author}{\bibfnamefont{K.~E.} \bibnamefont{Schmidt}},
  \bibinfo{journal}{in preparation}  (\bibinfo{year}{2003}).

\bibitem{krivine80}
\bibinfo{author}{\bibfnamefont{H.}~\bibnamefont{Krivine}},
  \bibinfo{author}{\bibfnamefont{J.}~\bibnamefont{Treiner}}, \bibnamefont{and}
  \bibinfo{author}{\bibfnamefont{O.}~\bibnamefont{Bohigas}},
  \bibinfo{journal}{Nuc. Phys. A} \textbf{\bibinfo{volume}{336}},
  \bibinfo{pages}{155} (\bibinfo{year}{1980}).

\bibitem{vautherin72}
\bibinfo{author}{\bibfnamefont{D.}~\bibnamefont{Vautherin}} \bibnamefont{and}
  \bibinfo{author}{\bibfnamefont{D.~M.} \bibnamefont{Brink}},
  \bibinfo{journal}{Phys. Rev C} \textbf{\bibinfo{volume}{5}},
  \bibinfo{pages}{626} (\bibinfo{year}{1972}).

\bibitem{ravenhall72}
\bibinfo{author}{\bibfnamefont{D.~G.} \bibnamefont{Ravenhall}},
  \bibinfo{author}{\bibfnamefont{C.~D.} \bibnamefont{Bennet}},
  \bibnamefont{and} \bibinfo{author}{\bibfnamefont{C.~J.}
  \bibnamefont{Pethick}}, \bibinfo{journal}{Phys. Rev. Lett.}
  \textbf{\bibinfo{volume}{28}}, \bibinfo{pages}{978} (\bibinfo{year}{1972}).

\bibitem{pandharipande89}
\bibinfo{author}{\bibfnamefont{V.~R.} \bibnamefont{Pandharipande}}
  \bibnamefont{and} \bibinfo{author}{\bibfnamefont{D.~H.}
  \bibnamefont{Ravenhall}}, in \emph{\bibinfo{booktitle}{Proceedings of the
  NATO Advanced Research Workshop on Nuclear Matter and Heavy Ion Collisions,
  Les Houches. NATO-ASI, Ser. B}}, edited by
  \bibinfo{editor}{\bibfnamefont{M.~S.} \bibnamefont{et~al.}}
  (\bibinfo{publisher}{Plenum}, \bibinfo{address}{New York},
  \bibinfo{year}{1989}), vol. \bibinfo{volume}{205}, p. \bibinfo{pages}{193}.

\end{thebibliography}

\end{document}